\documentclass[twocolumn]{aastex63}
\usepackage{graphicx}	
\usepackage{amsmath}
\usepackage{amssymb}	
\usepackage{color}
\usepackage{array}
\usepackage{bookmark}
\usepackage{subfigure}
\usepackage{booktabs}
\usepackage{threeparttable}

\begin{document}

\title{Variabilities driven by satellite black hole migration in AGN disks}

\author[0009-0008-9726-9431]{Jing-Tong Xing}
\affiliation{Department of Astronomy, Xiamen University, Xiamen, Fujian 361005, China}

\author[0000-0001-8678-6291]{Tong Liu}\thanks{E-mail: tongliu@xmu.edu.cn}
\affiliation{Department of Astronomy, Xiamen University, Xiamen, Fujian 361005, China}

\author[0000-0002-4448-0849]{Bao-Quan Huang}
\affiliation{College of Intelligent Manufacturing, Nanning University, Nanning, Guangxi 530299, China}

\author[0000-0002-0771-2153]{Mouyuan Sun}
\affiliation{Department of Astronomy, Xiamen University, Xiamen, Fujian 361005, China}

\begin{abstract}
The physical origin of active galactic nucleus (AGN) variability remains unclear. Here we propose that the magnetic reconnection induced by the migration of satellite black holes (sBHs) in the AGN disk can be a new plausible mechanism for AGN short-term variability. During the sBH migration, the co-moving plasmas surrounding the sBH could influence the large-scale magnetic field of the AGN disk and trigger the magnetic reconnections to contribute to AGN UV/optical variability. Meanwhile, high-magnetization plasmas are more likely to escape the disk and cause a secondary magnetic reconnection in the corona. For a $\sim 10^{2}-10^{3}~{M_\mathrm{\odot}}$ sBH in the inner regions of the disk surrounding a supermassive black hole with $\sim 10^{7}~{M_\mathrm{\odot}}$, the reconnection process occurred in the space out of the disk should produce X-ray emission, which can last $\sim 10^3-10^6~\rm s$ with the luminosity $\sim 10^{38}- 10^{42}~\rm{erg ~s^{-1}}$.
\end{abstract}


\section{Introduction} \label{sec: intro}

Active galactic nuclei (AGNs) are known to exhibit significant flux variations across multiple wavebands \cite[e.g.,][]{2024MNRAS.529.2877M}. UV/optical variability is believed to originate from geometrically thin, yet optically thick, accretion disks \cite[e.g.,][]{1973A&A....24..337S}, typically displaying small-amplitude, random fluctuations \cite[e.g.,][]{2021Sci...373..789B}. X-ray fluxes also show variability on kilo-second timescales \cite[e.g.,][]{2003MNRAS.345.1271V, 2013MNRAS.430L..49M}, with the amplitude of variability increasing with the accretion rate and anti-correlating with the mass of the supermassive black hole \cite[SMBH, e.g.,][]{2013ApJ...779..187K, 2016ApJ...819..154L}. In addition to stochastic variations, AGNs may also exhibit violent flares that significantly deviate from the baseline variability \cite[e.g.,][]{2010A&A...512A...1M}. However, it remains unclear whether these variations are all driven by a single physical mechanism. Various models have been proposed to explain AGN variability. These include UV/optical variability resulting from variable X-ray reprocessing \cite[e.g.,][]{1991A&A...249..344C, 1991ApJ...371..541K}, coronal heating and accretion disk reprocessing \cite[e.g.,][]{2020ApJ...891..178S}, UV/optical variability driven by global \cite[e.g.,][]{2008MNRAS.387L..41L, 2016MNRAS.462L..56L} or local \cite[e.g.,][]{1997MNRAS.292..679L} changes in the accretion rate, and X-ray variability due to coronal flares \cite[e.g.,][]{2004A&A...420....1C}. Other models include stochastic perturbations in the accretion disk \cite[e.g.,][]{2008ApJ...686..138F} or X-ray variability arising from a locally unstable, advection-dominated disk \cite[e.g.,][]{1996ApJ...464L.135M}. X-ray variability is generally thought to originate from the hot and optically thin corona in the inner regions of the AGN, which is theoretically linked to magnetohydrodynamic (MHD) instability and turbulence within the corona \cite[e.g.,][]{1991ApJ...380L..51H, 2002ApJ...572L.173L}.

The accretion disk around an SMBH contains many stars and compact objects that migrate due to density perturbations in the disk gas \cite[e.g.,][]{2016ApJ...819L..17B}. Similar to planetary migration in protoplanetary disks \cite[e.g.,][]{1996Icar..124...62P, 2010apf..book.....A}, compact objects typically migrate through the AGN disk mainly in one of two modes. When the mass ratio between a compact object and the central SMBH is small enough, such that the compact object migrates rapidly due to disk torque, the migration is referred to as Type I migration. This process has been studied in detail \cite[e.g.,][]{2010MNRAS.401.1950P, 2012MNRAS.425..460M}. If the compact object's mass is large enough, it can open a gap in the AGN disk \cite[e.g.,][]{1995MNRAS.277..758S, 2012MNRAS.425..460M}. This gap locks the compact object into the viscously driven accretion flow, causing it to migrate more slowly, a process known as Type II migration \cite[e.g.,][]{1986ApJ...309..846L, 1997Icar..126..261W, 2018ApJ...861..140K}.

\begin{figure*}
\centering
\includegraphics[width=0.6\textheight]{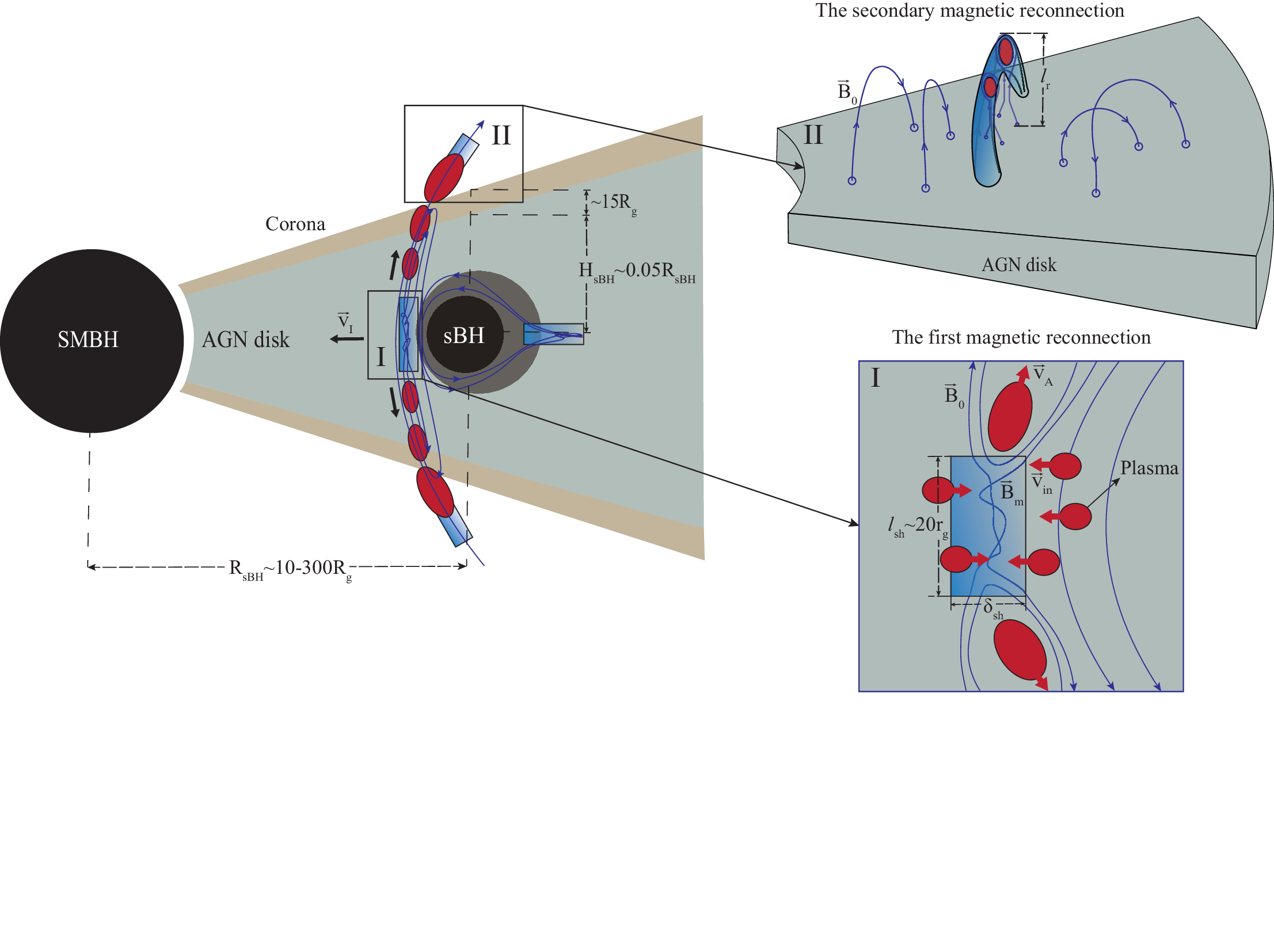}
\caption{Schematic diagram of the migration model. As an sBH migrates with $v_\mathrm{I}$ through the inner region of the AGN disk. The ram pressure of the surrounding plasma exceeds the magnetic field pressure, altering the topology of the magnetic field and causing the positive and negative magnetic fields to intertwine in the current sheets behind or trailing the sBH, eventually triggering the first magnetic reconnection within the disk. The current sheet behind the sBH is indicated by the boxed region (region I), where reconnection is likely to occur. The red ellipses in the picture represent plasma masses, and the blue boxes represent the current sheet. $l_\mathrm{sh}$ and $\delta_\mathrm{sh}$ respectively represent the length and width of the current sheet, while $v_\mathrm{in}$ and $v_\mathrm{A}$ are the inflow and outflow speeds of plasma in the current sheet. $B_\mathrm{0}$ and $B_\mathrm{m}$ are the magnetic field intensities within the initial and current sheet. This process effectively dissipates magnetic energy and accelerates the plasma. The ejected plasmoid expands and cools within the disk, and highly magnetized plasma can escape at the Alfven velocity, inducing the second magnetic reconnection (region II) in the AGN corona, which leads to X-ray variability. Additionally, plasma trapped in the disk heats the surrounding gas, contributing to UV/optical variability.}
\label{fig1}
\end{figure*}

Planetary migration is accompanied by planetary accretion, which affects the distribution of gas around the planet and its migration rate \cite[e.g.,][]{2017A&A...598A..80D, 2021ApJ...906...52L}. This effect is also relevant for compact objects in AGN disks. Their interactions with the disk may contribute to X-ray emission, and this contribution should be taken into account when studying AGN variability.

According to MHD simulation results, magnetic reconnection is believed to be common in black hole (BH) accretion disks and coronas \cite[e.g.,][]{2009MNRAS.395.2183Y, 2018ApJ...862L..25Z, 2023A&A...677A..67E}. Many current sheets developed due to turbulence and MHD instability provide sites for magnetic reconnection. Under ideal conditions, the magnetic field configuration breaks on the current sheet, creating a plasma chain (or magnetic island) with high magnetization \cite[e.g.,][]{2007PhPl...14j0703L, 2010PhRvL.105w5002U,2021JPlPh..87e9012R,2022ApJ...924L..32R}. A significant amount of energy from the magnetic field is dissipated through magnetic reconnection, accelerating the resulting plasma \cite[e.g.,][]{2009MNRAS.397.1153K, 2009ApJ...698.1570L}. The high magnetic field strength in the magnetically arrested disk provides an ideal environment for magnetic reconnection \cite[e.g.,][]{2003PASJ...55L..69N}, a property confirmed in M87 \cite[e.g.,][]{2019ApJ...875L...4E,2022ApJ...924..124Y}.

In this paper, we investigate a series of magnetic reconnection phenomena triggered by the migration of a satellite black hole (sBH) in the AGN disk. Section \ref{sec: Migration} shows the migration speeds under different modes and the conditions that trigger magnetic reconnection during migration. Section \ref{sec: Numerical analysis result} presents the numerical analysis results of magnetic reconnection, and the conclusions and discussion are provided in Section \ref{sec: Conclusion}.

\section{Migration} \label{sec: Migration}

The physical structure of the model is shown in Figure \ref{fig1}. We assume that the co-migration velocity of plasmas around the sBH reflects the effective migration velocity, which ``stamps'' the magnetic field lines, that is, the plasma's inertial motion imprints its trajectory onto the field lines. Distortion occurs once this dynamic imprinting overcomes the magnetic pressure. This stamp pressure affects the magnetic field topology, causing small-scale current sheets to be generated during migration. This, in turn, triggers magnetic reconnection, which accelerates the plasma. A portion of the plasma that successfully escapes from the disk continues to influence the magnetic field outside the disk, leading to secondary magnetic reconnection in the corona. Meanwhile, the rest of the plasma clumps trapped in the disk heat the surrounding gas, contributing to the thermal radiation of the AGN disk.

For an SMBH of mass $M_\mathrm{\bullet}=10^7~{M_\mathrm{\odot}}$ surrounded by a typical AGN disk with the viscous parameter $\alpha=0.01$ and the dimensionless accretion rate $\dot{m}=\dot{M}/\dot{M}_\mathrm{Edd}=0.15$, where $\dot{M}$ and $\dot{M}_{\rm Edd}$ are the mass accretion rate and Eddington accretion rate, respectively, we consider an sBH of mass $M_\mathrm{sBH}$ migrating in the AGN disk and assume its initial location is within the AGN disk migration trap \cite[$10-300~R_\mathrm{g}$, e.g.,][]{2016ApJ...819L..17B}, where $R_\mathrm{g}=2GM_\mathrm{\bullet}/c^{2}$ is the Schwarzschild radius of SMBH, $c$ is the speed of light.

\subsection{Migration speeds} \label{subsec: Migration}

Objects embedded in the disk experience different migration patterns with different masses. According to the simulations \citep[e.g.,][]{2018ApJ...861..140K}, for different migration modes, there is a transition condition between two migration types, that is
\begin{equation}
(\frac{M_\mathrm{sBH}}{M_\mathrm{\bullet}})_\mathrm{trans}=2.53 \times 10^{-4}(\frac{\alpha}{0.01})^{1/2}(\frac{h}{0.05})^{5/2},
\label{eq1}
\end{equation}
where the aspect ratio $h=H_\mathrm{sBH}/R_\mathrm{sBH}$, and $H_\mathrm{sBH}$ is the scale height of the disk where sBH is located at $R_\mathrm{sBH}$. When the ratio $q$ of the sBH mass to the SMBH is larger than $(M_\mathrm{sBH}/M_\mathrm{\bullet})_\mathrm{trans}$, the migration mode is in Type II; otherwise, it is the Type I migration.

For Type I migration, \cite{2010MNRAS.401.1950P} proposes that the torque can be described in terms of the linear Lindblad torque and nonlinear horseshoe torque. The migration velocity $v_\mathrm{I}$ and disk density $\rho_\mathrm{I}$ can be simplified to
\begin{equation}
v_\mathrm{I}=\frac{2\Gamma_\mathrm{c, hs}}{M_\mathrm{\bullet}R_\mathrm{sBH}\Omega_\mathrm{sBH}}\approx x_\mathrm{s}R_\mathrm{sBH}\Omega_\mathrm{sBH},
\label{eq2}
\end{equation}
where $\Gamma_\mathrm{c, hs}$ is the corotation torque in the horseshoe region \cite[e.g.,][]{2010MNRAS.401.1950P}, $\Omega_\mathrm{sBH}=(GM_\mathrm{\bullet}/R_{\rm sBH}^3)^{1/2}$ is the angular velocity of the sBH's orbit, and $x_{\rm s}\approx1.2\sqrt{q/h} $ is the half-width of the horseshoe shaped region \cite[e.g.,][]{2009MNRAS.394.2297P}; and
\begin{equation}
    \rho_\mathrm{I}=\left\{
\begin{aligned}
     \rho_\mathrm{0}(1-2\frac{\xi}{\hat{\gamma}}{x}),
 &&  {0<x<x_\mathrm{s},}\\
  \rho_\mathrm{0},  &  &\rm{otherwise,}
\end{aligned}
\right.
\label{eq3}
\end{equation}
where $x=(R-R_\mathrm{sBH})/R_\mathrm{sBH}$ is the distance of the disk gas from sBH in the radial direction, $\rho_\mathrm{0}$ is the initial density of the AGN disk, $\xi$ is the power-law index for the initial entropy curve, and $\hat{\gamma}$ is the adiabatic index.

The classical Type II migration model assumes that the gravitational torque of a massive compact object creates a deep gap in a narrow region, with the density within the gap being low enough for the Type I torque to become inefficient and no flows to cross the gap. Due to the large depth of the gap, the radial flow of gas through it ceases, causing the gas to be blocked and accumulate. The accumulated gas then exerts pressure on the compact object, driving it to move at a viscous speed \cite[e.g.,][]{1999MNRAS.307...79I, 2023ASPC..534..685P}. Under these assumptions, the migration density and velocity of sBH were resulted by Equations (39) and (40) in \cite{1995MNRAS.277..758S}, respectively.

While other types of migration and more complex torque-induced speed forms may exist, these are generally less stable and can even result in runaway behavior compared to the two types stated here \cite[e.g.,][]{2003ApJ...588..494M}.

\subsection{Condition of magnetic reconnection during migration} \label{subsec: reconnection translation}

High-resolution GRMHD simulations have observed the presence of highly magnetized regions near BHs, which provide a venue for the formation of current sheets \cite[e.g.,][]{2020MNRAS.495.1549N,2022MNRAS.513.4267N,2021JPlPh..87e9012R,2022ApJ...924L..32R}. These studies confirm that magnetic reconnection events near the event horizon generate plasma clumps, which can grow to several gravitational radii in size. As these clumps evolve and cool, they gain additional energy through continuous magnetic reconnection, enabling them to escape outward from the sBH vicinity. During this process, the clumps stretch, elongate, and sometimes fragment under significant shear forces. In the turbulent environment of sBH migration, current sheets form and dissipate rapidly, while fluid motion distorts magnetic field lines \cite[e.g.,][]{2022MNRAS.513.4267N}.

During migration, when the ram pressure of plasma $P_\mathrm{r}$ exceeds the magnetic pressure $P_\mathrm{0}$, the topology of the magnetic field lines can change dramatically, triggering magnetic reconnection. This process is similar to how the solar wind affects the magnetospheric structure of Earth \cite[e.g.,][]{1949srrs.book.....H, 1961PhRvL...6...47D, 2020JGRA..12525935H} and how the ``cosmic comb'' influences the magnetospheric structure of a pulsar to power fast radio bursts \cite[e.g.,][]{2017ApJ...836L..32Z}. We consider that the magnetic field is frozen in the disk plasma and transmitted to the SMBH through the inward motion of the plasma. The dynamics on the disk affect the value of the magnetic field, and the SMBH will swallow the plasma with low magnetic pressure. In contrast, the plasma with high magnetic pressure will be pushed back to the disk and continue to be modulated. Only when the magnetic pressure of the disk and the pressure of the inner region of the disk are balanced, does the magnetic field maintain the strength suitable for the inner region of the disk. For a disk with a stable accretion rate, the relation of the magnetic field and accretion rate at $R_\mathrm{sBH}$ can be expressed as $B_{0}=({2\dot{M}c}/{R_{\rm sBH}^{2}})^{1/2}$ \cite[e.g.,][]{1997MNRAS.292..887G,1999ASPC..190..173L,2017NewAR..79....1L}. The corresponding magnetic pressure can be calculated as $P_\mathrm{0}={B_\mathrm{0}^2}/{8\pi}$.

The magnetized plasmas around the sBH can effectively accelerate in the region where the magnetic field lines are highly curved and reconnect to release their energy. \cite{2017ApJ...836L..32Z} pointed that considering a non-relativistic and cold flow, the ram pressure of the plasma can be simplified as $P_\mathrm{r}\approx \rho_\mathrm{c} v^2$, where $\rho_\mathrm{c}$ is the co-migration plasma density. Assuming a local gas velocity of $v_\mathrm{I}$ near the sBH, the corresponding kinetic temperature is $T\approx m_\mathrm{p}v_\mathrm{I}^2/2k_\mathrm{B}\sim10^6~ \rm{K}$, where $m_{\rm p}$ is the proton mass, and $k_\mathrm{B}$ is the Boltzmann constant, sufficient to fully ionize hydrogen. In contrast, undisturbed disk gas at larger radii with $T\lesssim10^3~\rm{K}$ remains molecular, consistent with a typical AGN disk. The density of plasma flow is comparable to that of gas, and $P_\mathrm{r}$ under different migration modes can be expressed as
\begin{equation}
   P _\mathrm{r}=\left\{
\begin{aligned}
&\rho_\mathrm{0}(1-2\frac{\xi}{\hat{\gamma}}{x_\mathrm{s}})v_\mathrm{I}^2,
~~{{M_\odot}<M_\mathrm{sBH}<2.5\times10^{3}{M_\odot},}
 \\ &\rho_\mathrm{0}~
 (\frac{M_\mathrm{D}}{M_\mathrm{sBH}})^{\frac{3}{8}}(\frac{2}{3}\frac{R_\mathrm{sBH}}{\nu_\mathrm{0}})^{-2},~~\rm{otherwise,}
\end{aligned}
\right.
\label{eq4}
\end{equation}
where the viscosity is given by $\nu_\mathrm{0}=\alpha H^{2}\Omega_\mathrm{sBH} $ \cite[e.g.,][]{1973A&A....24..337S}, and $M_D=4\pi R_{\rm sBH}^2 \Sigma_{\rm sBH}$ is the disk mass with $\Sigma_{\rm sBH}$ being the surface density at $R_{\rm sBH}$. We consider that $x=x_\mathrm{s}$ to calculate the minimum disk density for Type I migration and take the constant $\xi=1.0$ and $\hat{\gamma}=5/3$ for the adiabatic disk \cite[e.g.,][]{1995MNRAS.277..758S}.

\begin{figure*}
\centering
\includegraphics[width=0.45\linewidth]{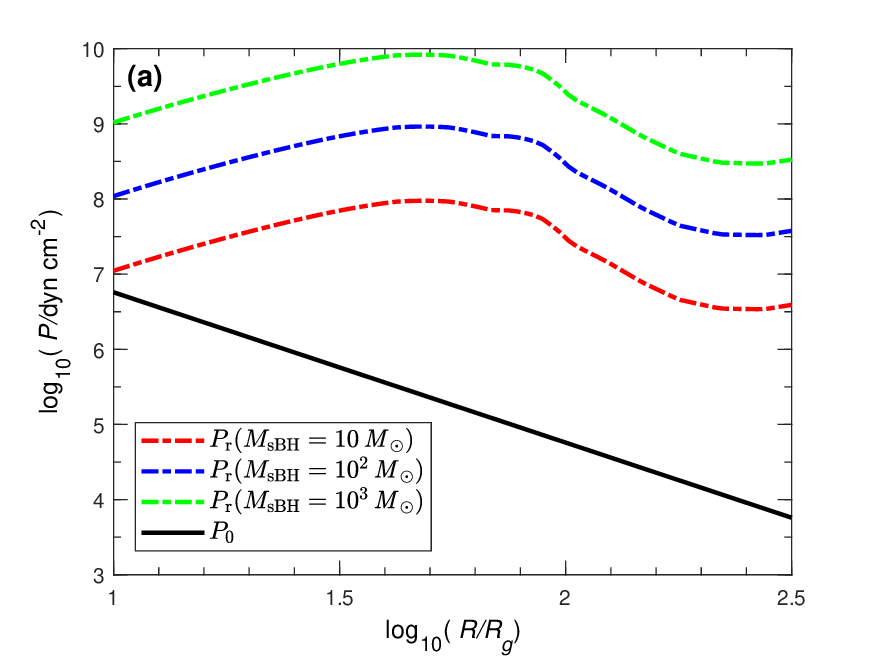}
\includegraphics[width=0.45\linewidth]{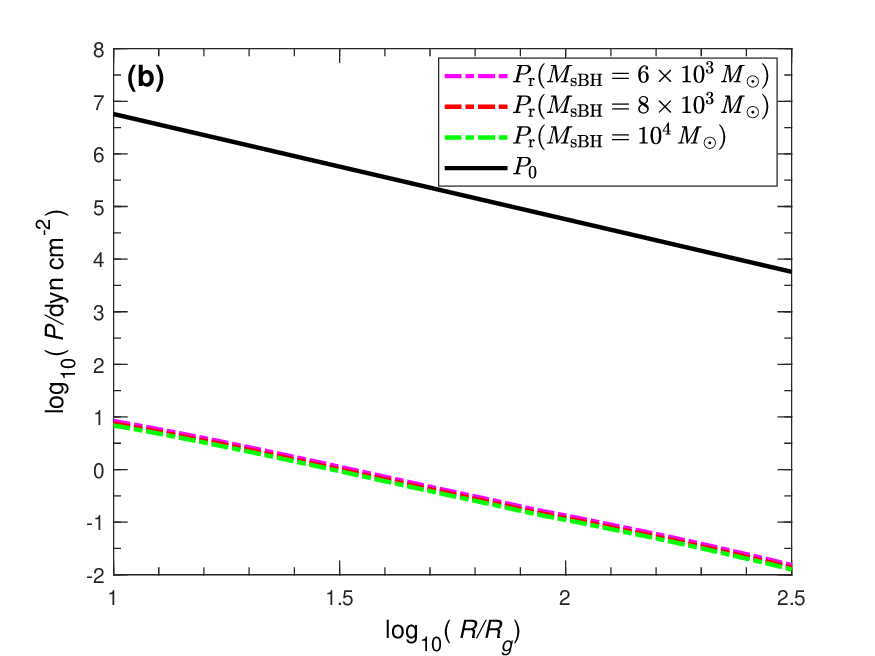}
\caption{The ram pressure of the plasma and the magnetic pressure in the AGN disk model \citep{2003MNRAS.341..501S} from $10 R_\mathrm{g}$ to $300 R_\mathrm{g}$ for $M_\mathrm{\bullet}=10^7M_\mathrm{\odot}$, $\dot{m}=\dot{M}/\dot{M}_\mathrm{Edd}=0.15$, and accretion efficiency $\epsilon=0.1$. (a) The ram pressure of Type I migration for $M_\mathrm{sBH}=10,~10^2$, and $10^3~M_\odot$ (corresponding to the red, blue, and green dashed lines), and the magnetic pressure (corresponding to the black solid line); (b) The same result of Type II migration for $M_\mathrm{sBH}=6\times10^3,~8\times10^3$, and $10^4~M_\odot$ (corresponding to the pink, red, and green dashed lines).}
\label{fig2}
\end{figure*}

Based on the above results, we calculate the AGN disk model \citep{2003MNRAS.341..501S} to compare $P_\mathrm{r}$ with $P_\mathrm{0}$ during migration. The results are shown in Figure \ref{fig2}, at the migration trap considered in our model ($10-300~R_\mathrm{g}$), the sBH undergoing Type I migration are more prone to trigger magnetic reconnection events in the disk, while massive sBH undergoing Type II migration are less likely to do so. Based on this finding, we only analyze the case of Type I migration in the subsequent calculations.

\section{Magnetic reconnection} \label{sec: Numerical analysis result}

We propose that during sBH migration, two types of magnetic reconnection events are initiated in the disk and the corona, respectively. We select a set of typical parameter analyses to calculate the process of magnetic reconnection. Considering an sBH with mass $M_\mathrm{sBH}=2\times10^2~{M_\mathrm{\odot}}$ migrates in the AGN disk, the speed of migration and the magnetic field strength are
\begin{equation}
\begin{aligned}
v_\mathrm{I}=5.09\times10^{7}{m_\mathrm{sBH,2}}^{1/2}h_\mathrm{0.05}^{-1/2}r_\mathrm{sBH,2}^{-1/2}~\rm {cm~s^{-1}},
\label{eq5}
\end{aligned}
\end{equation}
and
\begin{equation}
B_\mathrm{0}=1.20\times10^{3}\dot{m}_\mathrm{0.15}^{1/2}r_\mathrm{sBH,2}^{-1}~\rm {G},
\label{eq6}
\end{equation}
respectively, where $m_\mathrm{sBH,2}=M_\mathrm{sBH}/(2\times10^2~{M_\mathrm{\odot}})$, $h_\mathrm{0.05}=h/0.05$, $r_\mathrm{sBH,2}= R_\mathrm{sBH}/(10^2~R_\mathrm{g})$, and $\dot{m}_\mathrm{0.15}=\dot{M}/(0.15~\dot{M}_\mathrm{Edd})$.

\subsection{Magnetic reconnection in AGN disks} \label{subsec: First Rec}

The sBH with Type I migration may trigger magnetic reconnection due to the tearing instability \cite[e.g.,][]{2007ApJ...670..702Z}. Due to turbulence affecting the toroidal magnetic field, current sheets form throughout the ring. Magnetic reconnection proceeds at a significantly high rate within the current sheet, where magnetic field lines are strongly sheared and twisted by the surrounding plasma flow. The reconnection of magnetic flux tubes in this region efficiently converts magnetic energy into thermal energy, heating the plasma to near-relativistic temperatures and enabling it to overcome gravitational confinement \cite[e.g.,][]{2014ApJ...783L..21S}. As shown in the lower right subfigure of Figure \ref{fig1}, the blue region is referred to as the current sheet, where magnetic reconnection occurs as the bound plasma streams migrate with the sBH and the magnetic field lines within the AGN disk bend. When two magnetic field lines of opposite directions and the surrounding plasma enters the current sheet, the topology of the magnetic field changes, and the oppositely directed field lines effectively break and reconnect. Within this region, high magnetization and turbulent conditions drive efficient particle acceleration, generating a relativistic electron population that emits synchrotron radiation \cite[e.g.,][]{2020JGRA..12525935H}. Theoretical studies generally also show that the initial structure of the magnetic field breaks within the current sheet, creating plasma chains or magnetic islands \cite[e.g.,][]{2007PhPl...14j0703L, 2010PhRvL.105w5002U}.

Due to the ram pressure of the plasma, the magnetic flux will be accumulated in the current sheet, and amplify the local magnetic field intensity to provide fuel for the reconnection. We assume the typical length scale of a current sheet $l_\mathrm{sh}=20 r_\mathrm{g}$ \cite[e.g.,][]{2021JPlPh..87e9012R,2022ApJ...924L..32R}, and width scale $\delta_\mathrm{sh}={g}l_\mathrm{sh}$, where $r_\mathrm{g}=2GM_\mathrm{sBH}/c^2$ is the Schwarzschild radius of the sBH and ${g}$ is the current sheet geometric index. The gas-to-magnetic-pressure for the plasma flows is $\beta_\mathrm{p}= P_\mathrm{r}/P_\mathrm{m}$. The low $\beta_\mathrm{p}$ gaps form during the static process before the magnetic flux eruption and show obvious instability at the gap boundaries \cite[e.g.,][]{2022MNRAS.513.4267N,2022ApJ...924L..32R}. Thin current sheets are formed in the magnetosphere. Here we assume $\beta_\mathrm{p}\sim 0.01$ at the current sheet, corresponding to a magnetic pressure $100$ times greater than the gas pressure. This value is characteristic of strongly magnetized environments such as AGN corona, the base of relativistic jets, and highly magnetized disk surface layers. For instance, particle-in-cell (PIC) simulations by \cite{2024GeoRL..5112126Y} show that low-$\beta_\mathrm{p}$ facilitates sheet pinching to kinetic scales and onset of reconnection, while MHD models by \cite{2018ApJ...852...95N} confirm fast reconnection in $\beta_\mathrm{p} \ll 0.01$ environments even under partially ionized, kilogauss-field conditions. Because of the magnetic flux conservation, the large-scale magnetic field of the accretion disk is compressed into the scale of the current sheet, i.e., $B_\mathrm{m}l_\mathrm{sh}\delta_\mathrm{sh}=B_\mathrm{0}v_\mathrm{I}t_\mathrm{c}2H$ and reconnection at $P_\mathrm{m}\sim \beta_\mathrm{p}^{-1} P_\mathrm{r}={B_\mathrm{m}^2}/{8\pi}$, where $B_\mathrm{m}$ is the magnetic field after accumulation. The compression timescale can be estimated as $t_\mathrm{c}\approx (8\pi \beta_\mathrm{p}^{-1} P_\mathrm{r})^{1/2}l_\mathrm{sh}\delta_\mathrm{sh}/(B_\mathrm{0}v_\mathrm{I}2H)$.

\cite{2024PhRvD.110f3003C} presents that the local reconnection rate is related to the current sheet geometric index and the magnetization. In the case of high local-magnetization limit ($\sigma\gg1$), the reconnection rate is
\begin{equation}
  {\mathcal{R}}_\mathrm{rec} \simeq {g} \frac{1-{g}^{2}}{1+{g}^{2}} \sqrt{\frac{\left(1-{g}^{2}\right)^{3} \sigma}{\left(1+{g}^{2}\right)^{2}+\left(1-{g}^{2}\right)^{3} \sigma}},
  \label{eq7}
\end{equation}
where $\sigma$ is the magnetization, which is defined as the ratio of the magnetic energy to the inertial mass energy of the particles \cite[e.g.,][]{2022A&A...663A.169E, 2023A&A...677A..67E}. In addition, the reconnection rate can also be expressed as the ratio of the inflow velocity of the plasma to the local Alfven velocity ${\mathcal{R}{}}_\mathrm{rec}=v_\mathrm{in}/v_\mathrm{A}\sim 0.1$ \cite[e.g.,][]{2013MNRAS.431..355G,2025arXiv250602101S}, note that $v_\mathrm{A}=\sqrt{\sigma/(1+\sigma)} c$ at high magnetization \cite[e.g.,][]{2013MNRAS.431..355G,2025arXiv250602101S}. The reconnection time within the current sheet can be obtained as $t_\mathrm{r,in}\approx \delta_\mathrm{sh}/v_\mathrm{in}$. As Figure \ref{fig1} shows, the local magnetic field strength ahead of the sBH is enhanced, leading to a larger Alfven speed, a thinner current sheet, and a stronger reconnection electric field $E_\mathrm{rec}\sim v_\mathrm{in}B\sim \mathcal{R}v_\mathrm{A}B$. In contrast, the trailing region may feature weaker magnetic fields and reduce reconnection efficiency. Additionally, the plasma accelerated in the trailing current sheet is more difficult to escape from the disk. These differences justify our focus on the leading current sheet.

As alternating polar flux tubes interact and reconnect, the highly magnetized plasma produced in the current sheet region may escape rather than fall back into the sBH. As it travels outward and cools, this process significantly increases the disk temperature \citep[e.g.,][]{2022MNRAS.513.4267N}. For highly magnetized plasma, the speed approaches the speed of light, and there is a high probability that it can escape the disk. We consider cooling via Coulomb collisions between relativistic electrons and protons to determine whether rapidly cooling plasma cannot escape. The cooling timescale can be estimated as $t_\mathrm{cool} \approx \frac{3}{4}{(\frac{m_\mathrm{e}}{2\pi})}^{1/2}(\gamma_\mathrm{e,out} m_\mathrm{e} c^{2})^{3/2}/({\rm e}^4 n_\mathrm{p} \mathrm{ln} \Lambda)$ \cite[e.g.,][]{1962pfig.book.....S,2012ApJ...754...92C}, where $m_\mathrm{e}$ is the electronic mass, $\gamma_\mathrm{e,out}$ is the Lorentz factor of the electrons in the escape plasma, $\rm e$ is the electronic charge, $n_\mathrm{e}=n_\mathrm{p} \sim \rho_\mathrm{0}/m_{\rm p}$ is the number density of electrons and protons, and the Coulomb logarithm is $\mathrm{ln} \Lambda\sim 10$. {To ensure that magnetic reconnection can effectively occur during the migration process and that accelerated relativistic particles can escape from the disk, the plasma needs to meet $t_\mathrm{cool}\gtrsim t_\mathrm{dyn}\sim H/v_\mathrm{A}\gtrsim t_\mathrm{c}\gtrsim t_\mathrm{r,in}$, these timescales are all determined by the magnetization intensity $\sigma$. We assume the minimum magnetization of plasmas as the same as the equipartition argument gives $\sigma_\mathrm{min}=3$ for the electron in the plasmoid \cite[e.g.,][]{2013MNRAS.431..355G}, and the maximum magnetization as $\sigma_\mathrm{max}=25$ \cite[e.g.,][]{2021JPlPh..87e9012R}, the corresponding electron Lorentz factor is $\gamma_\mathrm{e,out}\sim 10^3\sigma$  \cite[e.g.,][]{2013MNRAS.431..355G}. The comparison results of the four timescales are shown in the Figure \ref{fig3}(a). The results show that for the case of $\sigma \leq 3$, $t_\mathrm{cool}$ is much smaller than $t_\mathrm{dyn}$, that is, most of plasmas are cooled within the disk.

The simulation results by \cite{2020MNRAS.495.1549N} show that the plasma may escape when the magnetization parameter $\sigma\ge3$ \cite[also see e.g.,][]{2017ApJ...843...21L,2018ApJ...868..101B}. This result is consistent with the magnetization parameters required by the four characteristic timescales. However, this is not always the case, and parts of the plasma will be advected and accreted by the sBH, or bound to the disk. These plasmas contain a large number of relativistic particles, which may be the origin of flares on the AGN disk \cite[e.g.,][]{2013MNRAS.431..355G,2017ApJ...843...21L}. Since these plasma spheres are generally thought to be filled with high-energy particles, the reconnection-accelerated plasma that does not break through may contribute to the disk's thermal radiation. Let's consider sBHs uniformly distributed in a two-dimensional plane $N(R)$, and the luminosity released $L_\mathrm{th}$ by magnetic reconnection is converted into the thermal energy of the disk. The total luminosity due to magnetic reconnection driven by sBHs migration at radius  $R_\mathrm{sBH}$ can be given as $L_\mathrm{sBH}^{+}=N(R)L_\mathrm{th}$. The distribution of sBHs, i.e., \cite[e.g.,][]{2024ApJ...966L...9Z}
\begin{equation}
N(R)=\frac{2 R_\mathrm{sBH} \Delta R N_\mathrm{sBH} }{(R_\mathrm{out}^{2}-R_\mathrm{in}^{2})},
\label{eq8}
\end{equation}
where $N_\mathrm{sBH}$ is the total number of sBHs. The kinetic energy of the plasma will be converted into heat energy in the disk, thereby heating the disk. \cite{1993ApJ...409..592A} pointed out that the AGN accretion disk could capture stars from nuclear star clusters at a rate of $10^{-3}-10^{-4}~\rm{yr^{-1}}$, the number of sBHs that can be captured within a typical AGN lifetime was $10^3-10^5$ \cite[e.g.,][]{2024ApJ...966L...9Z}. We assume that $N_\mathrm{sBH}=10^3$ in a region ($R_\mathrm{out}=300~R_\mathrm{g}$, $R_\mathrm{in}=10~R_\mathrm{g}$) and consider all the magnetic field energy during $t_\mathrm{cool}\sim t_\mathrm{dyn}$ was given to the thermal radiation. The magnetic energy released within a single thin current sheet is $E_\mathrm{r}=B_\mathrm{m}^2V_\mathrm{sheet}/8\pi$, where $V_\mathrm{sheet}=(\Omega_\mathrm{sBH}t_\mathrm{r,in}R_\mathrm{sBH})l_\mathrm{sh}\delta_\mathrm{sh}$ since $\Omega_\mathrm{sBH} t_\mathrm{r,in}<2 \pi$ \cite[e.g.,][]{2018ApJ...868...19W,2019ApJ...873L..13D}. Considering that migration is a dynamic process, multiple similar small current sheets may be generated during the process. Within $t_\mathrm{dyn}$, the number of sheets that can be generated is $N_\mathrm{sheet}\sim  t_\mathrm{dyn}/t_\mathrm{c}$,  then the thermal luminosity is
\begin{equation}
\begin{aligned}
&L_\mathrm{th}\approx\frac{N_\mathrm{sheet}B_\mathrm{m}^{2}V_\mathrm{sheet}}{8\pi t_\mathrm{r,in}}\\&=2.22\times10^{40}m_\mathrm{sBH,2}m_\mathrm{\bullet,7}^{1/2}\dot{m}_\mathrm{0.15}^{1/2}h_\mathrm{0.05}r_\mathrm{sBH,2}^{-1/2}\rho_\mathrm{-8}^{1/2}\beta_\mathrm{-2}^{-1/2}~\rm{erg~s^{-1}},
\end{aligned}
\label{eq9}
\end{equation}
where $m_\mathrm{\bullet,7}=M_\mathrm{\bullet}/(10^7~{M_\mathrm{\odot}})$ is the dimensionless SMBH mass, $\rho_\mathrm{-8}=\rho_\mathrm{I}/(10^{-8} \rm{g~cm^{-3}})$ is the dimensionless disk density, and $\beta_\mathrm{-2}=\beta_\mathrm{p}/(10^{-2})$. This result corresponds to the case where the magnetization is $\sigma_\mathrm{min}$. Figure \ref{fig3}(b) shows that $Q_\mathrm{sBH}^{+}/Q_\mathrm{vis}^{+}>1$ at $R_\mathrm{sBH}\approx 100~R_\mathrm{g}$ when $\sigma_\mathrm{min}<\sigma<\sigma_\mathrm{max}$ is satisfied, where $Q_\mathrm{sBH}^{+}=L_\mathrm{sBH}^{+}/(2 \pi R_\mathrm{sBH}\Delta R)$ is the heating rate per unit area for the magnetic reconnection driven by sBHs migration in the disk, $Q_\mathrm{vis}^{+}$ represents the disk viscous heat production, i.e., $Q_\mathrm{vis}^{+}=(3GM_\mathrm{\bullet}\dot{M} f_\mathrm{r})/(8\pi R_\mathrm{sBH}^{3})$ and $f_\mathrm{r}=1-(3R_\mathrm{g}/R_\mathrm{sBH})^{1/2}$ \cite[e.g.,][]{2024ApJ...966L...9Z}. The results indicate that for high magnetization limits $\sigma \gg 1$, a higher plasma magnetization leads to a lower fraction of kinetic energy being converted into thermal energy within the disk, and a correspondingly higher probability of energy escape. This is due to the reconnection layer ($l_\mathrm{sh}\delta_\mathrm{sh}$) becoming thinner, and dynamic timescales ($t_\mathrm{dyn}$) shortens. Together, these effects reduce the efficiency of energy dissipation into heat.

Assuming that sBHs are bound to a certain radius on the AGN disk due to migration traps, the migration of sBHs may affect the effective temperature at that radius, thus dominating the AGN UV/optical variability. When the sBH is trapped in a migration trap, the accumulation and disturbance of magnetic flux derived from migration movement will become relatively inactive, thereby reducing the occurrence frequency of magnetic reconnection events and heating efficiency. It should be noted that in real scenarios, the magnetic field becomes stronger over time during the dynamic period, which suggests that the scale of the current sheet may also be increasing. Our total scale of the sheets is equivalent to taking the average value for calculation.

\begin{figure*}
\centering
\includegraphics[width=0.45\linewidth]{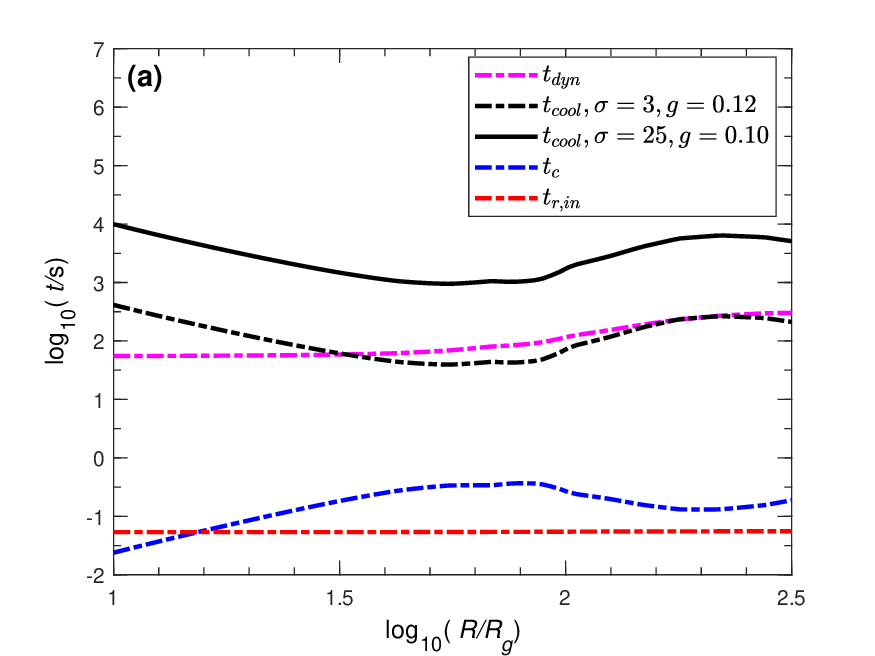}
\includegraphics[width=0.45\linewidth]{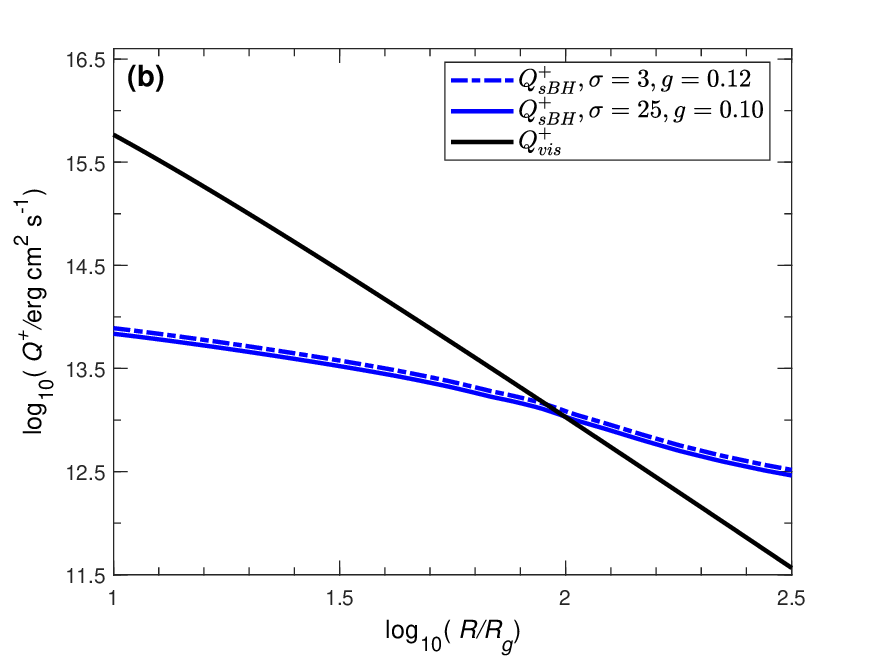}
\caption{(a) The dynamic timescale, cooling timescale, and compression timescale of the plasma from $10 R_\mathrm{g}$ to $300 R_\mathrm{g}$, as well as the magnetic reconnection timescale triggered within the current sheet. (b) The sBH heating rate and the disk viscous heating rate correspond to the blue and black lines. Solid and dashed blue lines represent the different magnetizations and geometric indices.}
\label{fig3}
\end{figure*}

\subsection{Magnetic reconnection in AGN coronas} \label{subsec: Second Rec}

The accelerated plasma carries a large number of relativistic particles, which will greatly enhance the magnetic turbulence in the AGN corona, resulting in secondary magnetic reconnection. However, the magnetic field in the accretion disk is usually turbulent, and the plasma does not escape strictly perpendicular to the disk, but is likely to be trapped inside the disk. The situation described here is more likely to occur in the AGN disk with low luminosity and high magnetic field when the plasma escapes the AGN disk along the magnetic field lines.

The AGN corona is thought to be similar to the solar corona in that it contains hot, quasi-spherical plasma, which contributes to the power-law component of the X-ray spectrum of the accreting BH \cite[e.g.,][]{1976SvAL....2..191B, 1977ApJ...218..247L}. Due to the interchange instability, the magnetic field is buoyant and rises into the corona, where it dominates the magnetic structure \cite[e.g.,][]{1979ApJ...229..318G, 1992MNRAS.259..604T}. We propose that most of the magnetic field in the AGN corona exists in the form of magnetic rings or flux ropes \cite[e.g.,][]{2008ApJ...682..608U}. These rings represent closed magnetic fields (as shown in the upper right subfigure of Figure \ref{fig1}, where the foot points of the rings are anchored at different radii on the disk surface). Magnetic rings floating on the disk surface can be twisted and stretched by differential Kepler rotation and turbulence in the accretion flow, causing the rings to expand \cite[e.g.,][]{2008ApJ...682..608U}. During torsion and expansion, the rings may undergo internal destruction due to MHD instabilities, or they may reconnect with other rings, releasing magnetic energy stored in the AGN corona \cite[e.g.,][]{2004ApJ...606.1083H, 2008ApJ...688..555G, 2009MNRAS.395.2183Y}.

The plasma flow and magnetic current sheet mainly determine the variation of the magnetic field structure over time, which can be expressed as
\begin{equation}
\frac{\partial \vec{B}}{\partial t}=\nabla \times(\vec{v} \times \vec{B})-\nabla \times(\eta \nabla \times \vec{B}),
    \label{eq10}
\end{equation}
where $\eta$ is the resistivity, $\vec{v}$ is the velocity of the plasma. As a large amount of plasma carrying relativistic particles enters the corona environment, the rapid increase of the first term on the right side of Equation (\ref{eq10}) sign causes the magnetic ring to be locally compressed and becomes strongly twisted, and the current density increases in a narrow area, forming a current sheet \cite[e.g.,][]{2002ApJ...564L..53L}. Both the current and magnetic field within and outside of the flux rope satisfy the following conditions $\vec{J}=\frac{c}{4 \pi} \nabla \times \vec{B}$ \cite[e.g.,][]{2000JGR...105.2375L}.

There are two scenarios for the interaction of plasma with the magnetic field in the corona: (i) The plasma flows along the direction of the magnetic field without forming magnetic reconnection, and the magnetic field lines flow with the fluid; (ii) when plasma is vertically or obliquely cut into the magnetic field region, it may compress and shear the magnetic field lines, resulting in magnetic field distortion. The inherent assumption of our model is situation (ii). To quantify whether the spatial structure of the magnetic field can be compressed by the plasma escaping at high speed, we compared the dynamic pressure of the plasma with the magnetic pressure of the corona magnetic ring $P_\mathrm{kin}\sim1/2 \rho_\mathrm{p}v_\mathrm{A}^2(1-t_\mathrm{dyn}/t_\mathrm{cool,25})\sim10^{8}~{\rm{dyn~cm^{-2}}}\gg P_\mathrm{0}$, where $\rho_\mathrm{p}\approx N_\mathrm{sheet}B_\mathrm{m}^{2}/({4\pi c^{2}\gamma_\mathrm{e,out}\sigma_\mathrm{max}})$ is the escape plasma density with high magnetization \cite[e.g.][]{2022A&A...663A.169E, 2023A&A...677A..67E}. The magnetic ring in the corona will be compressed by the escape plasmas and trigger the second reconnection outside the disk, as shown in the upper right subfigure of Figure \ref{fig1} \cite[e.g.,][]{2002ApJ...564L..53L,2009MNRAS.395.2183Y}.

To estimate the shortest migration timescale for the sBH position $R_\mathrm{sBH}$, we consider the inner region ($R_\mathrm{mig}\sim 20~R_\mathrm{g}$) where magnetic reconnection due to migration no longer dominates the emission of radiation. Notably, there are still certain mechanisms capable of triggering magnetic reconnection in the disk, such as MHD instability \cite[e.g.,][]{2020MNRAS.495.1549N} or magnetospheric collisions resulting from rotation \cite[e.g.,][]{2022A&A...663A.169E}. Still, these mechanisms need to be triggered in the vicinity of SMBH. The migration time at $R_\mathrm{sBH}=100~R_\mathrm{g}$ can be estimated by
\begin{equation}
\begin{aligned}
&t_\mathrm{m}=\frac{R_\mathrm{m}}{v_\mathrm{I}}=\frac{R_\mathrm{sBH}-R_\mathrm{mig}}{v_\mathrm{I}}
\\&\approx 4.66\times 10^{6}m_\mathrm{sBH,2}^{-1/2}h_\mathrm{0.05}^{1/2}r_\mathrm{sBH,2}^{1/2} r_\mathrm{m,80}~\rm s,
\end{aligned}
\label{eq11}
\end{equation}
where $r_\mathrm{m,80}=R_\mathrm{m}/(80~R_\mathrm{g})$ is migration distance. In fact, the migration timescale should be $\gg t_\mathrm{m}$ at $R_\mathrm{sBH}$ \cite[e.g.,][]{2012MNRAS.425..460M}, here we just give a lower bound. For the escape plasmas with magnetization $\sigma_\mathrm{min}$, the cooling time in the corona can be estimated $t_\mathrm{cool}(\rho_\mathrm{p})\sim\frac{3}{4}{(\frac{m_\mathrm{e}}{2\pi})}^{1/2}(\gamma_\mathrm{e,out} m_\mathrm{e} c^{2})^{3/2}/({\rm e}^4 (\rho_\mathrm{p}/m_\mathrm{p}) \mathrm{ln} \Lambda)$. We assume that the velocity of the escaping plasma is cooled and decelerated by the disk to $\sim \chi v_\mathrm{A}$, where $\chi>(P_\mathrm{0}/\rho_\mathrm{p})^{1/2}/c\sim0.001$ should be satisfied. The scale height of the current sheet in the corona is $l_\mathrm{r}\approx\chi v_\mathrm{A}t_\mathrm{cool}(\rho_\mathrm{p})$, and for the escape plasmas with magnetization $\sigma_\mathrm{max}$, the height of the current sheet needs to compare to the corona height. Based on the restriction of AGN corona height by \cite{2020NatAs...4..597A}, we consider the max height of the sheet out of the disk to be $l_\mathrm{r,max}\sim 15~R_\mathrm{g}$, and the width is $\Delta\sim{g}l_\mathrm{r}$. Hence, the duration of the magnetic reconnection is
\begin{equation}
\begin{aligned}
&t_\mathrm{r,out}=\frac{l_\mathrm{r}}{\mathcal{R}_\mathrm{rec} v_\mathrm{A}}
\\& \approx 2.26\times 10^{3}\mathcal{R}_\mathrm{0.1}^{-1} \chi_\mathrm{-3}m_\mathrm{sBH,2}\dot{m}_\mathrm{0.15}^{-1/2}h_\mathrm{0.05}^{-1}\rho_\mathrm{-8}^{-1/2}\beta_\mathrm{-2}^{1/2} ~\rm s,
\end{aligned}
\label{eq12}
\end{equation}
where $\chi_\mathrm{-3}=\chi/(10^{-3})$, $\mathcal{R}_\mathrm{0.1}=\mathcal{R}_\mathrm{rec}/0.1$ and $v_\mathrm{A}=\sqrt{\sigma_\mathrm{min}/(1+\sigma_\mathrm{min})}c$. It can be seen that the migration time $t_\mathrm{m} \gg t_\mathrm{r,out}$, which indicates that the sBH is likely to cause catastrophic magnetic reconnection during the migration process, resulting in continuous, longer-lasting flares.

The total magnetic energy can be calculated as ${B_\mathrm{0}^{2}V_\mathrm{r}}/{8\pi}$. The secondary magnetic reconnection luminosity in the corona caused by a single sBH is
\begin{equation}
\begin{aligned}
&L_\mathrm{r}=\frac{B_\mathrm{0}^{2}V_\mathrm{r}}{8\pi t_\mathrm{r,out}}\approx \frac{B_\mathrm{0}^{2}(\Omega_\mathrm{sBH}t_\mathrm{r,out}R_\mathrm{sBH}l_\mathrm{r}\Delta)}{8\pi t_\mathrm{r,out}}\\&=5.01\times10^{38}\chi_\mathrm{-3}^2m_\mathrm{sBH,2}^2m_\mathrm{\bullet,7}^{1/2}h_\mathrm{0.05}^{-2}r_\mathrm{sBH,2}^{-5/2}\rho_\mathrm{-8}^{-1}\beta_\mathrm{-2}~\rm{erg~s^{-1}}.
\end{aligned}
\label{eq13}
\end{equation}
Figure \ref{fig2}(a) shows that $P_\mathrm{r}\gg P_\mathrm{0}$ is still satisfied at less than $20~R_\mathrm{g}$, which means that powerful magnetic reconnection may still occur in the inner region of the disk. If we consider that there is an intermediate-mass BH in the disk region undergoing Type I migration, and the height of the current sheet is $l_\mathrm{r,max}$, the luminosity released during the migration is $L_\mathrm{max}\sim 7.57\times10^{42}~\rm{erg~s^{-1}}$ for $R_\mathrm{sBH} = 10~R_\mathrm{g}$. This luminosity can be comparable to the AGN X-ray luminosity. Based on the $L_\mathrm{2 \rm keV}-L_\mathrm{2500}$ relation $(\log L_\mathrm{2keV}=\omega \log L_\mathrm{2500}+\beta)$ with a slope $\omega$ of $\sim$ 0.6 and intercept $\beta$ of $\sim$ 7 \cite[e.g.,][]{2016ApJ...819..154L}, we can estimate the AGN X-ray luminosity of $\sim 4.98\times10^{42} ~\rm {erg~s^{-1}}$ for an accretion rate of $\sim 0.15~\dot{M}_\mathrm{Edd}$ by assuming that $L_\mathrm{2500}$ is consistent with the Eddington luminosity. Hence, our work presents a new mechanism to account for short-term X-ray variability in AGNs. It should be noted that we do not consider the magnetic amplification due to the flux accumulation, so that the strength of the magnetic field at reconnection will be stronger than the above estimation.

\section{Conclusions and discussion} \label{sec: Conclusion}

In this paper, we have investigated the AGN variability caused by the sBH migration. For objects migrating in the AGN disk that are subjected to torques caused by perturbations in disk density, \cite{2016ApJ...819L..17B} shows that migrating objects in different directions will meet due to changes in the sign of the torques, forming a migration trap. In the migration trap, massive sBHs can be formed efficiently. Our model suggests that as the sBH migrates through the disk, interactions with surrounding plasma clumps disrupt the magnetic field geometry, triggering intense magnetic reconnection. This process generates highly magnetized plasmas and plasma chains, rich in relativistic particles, which supply the energy responsible for the AGN disk's thermal radiation and X-ray variability. The main conclusions are as follows:
\begin{enumerate}
\item We demonstrate that the co-moving plasmas surrounding the sBH are more likely to influence the large-scale magnetic field during Type I migration. This interaction drives the initial magnetic reconnection within the AGN disk.
\item For the first magnetic reconnection, we consider the presence of $10^3$ sBH in the region between  $10~R_\mathrm{g}$ and $300~R_\mathrm{g}$. By comparing the heating rates of sBHs at various radii and the viscous heat generation of the disk, we consider that the magnetic reconnection effect during the migration process can affect the effective temperature of the disk, and contribute to UV/optical variability.
\item For high-magnetization plasmas that satisfy $t_\mathrm{cool}\gtrsim t_\mathrm{dyn}\gtrsim t_\mathrm{c}\gtrsim t_\mathrm{r,in}$, it is more likely to escape the disk environment and enter the corona. In addition, the comparison of the dynamic pressure and magnetic pressure of the escaping plasma also indicates that large-scale magnetic fields will undergo secondary magnetic reconnection outside the disk.
\item For the second magnetic reconnection, our results show that for a $\sim 10^{2}-10^{3}~{M_\mathrm{\odot}}$ sBH (including its binding gas) in the inner regions of the disk surrounding an SMBH with $\sim 10^{7}~{M_\mathrm{\odot}}$, the reconnection process occurs in the space out of the disk can produce observable X-ray variability superposed on the AGN light curve, which can last $\sim 10^3-10^6~\rm s$ with the luminosity $\sim 10^{38}- 10^{42}~\rm{erg ~s^{-1}}$.
\end{enumerate}

We note that most of the simulations referenced here model optically thin, geometrically thick accretion disks, which differ from the optically thick, geometrically thin disks relevant for many AGN. However, magnetic field evolution and reconnection are also expected in thin-disk simulations \citep[e.g.,][]{2016MNRAS.462..636A,2022ApJ...935L...1L,2024MNRAS.527.1424S,2025arXiv250602289Z}.

The sBH spin may also trigger magnetic reconnection in plasma flows \cite[e.g.,][]{2022A&A...663A.169E,2023A&A...677A..67E}. In our model, the rotational energy of the sBH was not taken into account, which means that the magnetic energy released in practice may be greater. In addition, the spin of sBH may also bring about relativistic jets \cite[e.g,][]{1977MNRAS.179..433B}, and the large amount of plasma driven by the jets might also be one of the factors triggering secondary magnetic reconnection in the corona.

Our results are independent of the sBH accretion rate, and future modifications to this formula will be necessary. Additionally, considering that other compact objects, such as white dwarfs and neutron stars, also migrate in AGN disks, the migration of a single low-mass compact object is unlikely to significantly affect the large-scale magnetic field structure of the AGN disk. It has been suggested that massive nuclear cluster objects (NCOs), which consist of stars and compact objects, surround SMBHs and migrate and merge within the disk \cite[e.g.,][]{2012MNRAS.425..460M}. We believe that binary accretion systems within these NCOs could produce effects similar to those arising from the migration of massive sBHs.

\acknowledgments
We thank anonymous referee for very helpful suggestions and comments, and Yun-Feng Wei, Shuying Zhou, Jiao-Zhen She, and Xiao-Yan Li for the helpful discussion. This work was supported by the National Key R\&D Program of China under grants 2023YFA1607902 and 2023YFA1607903, the National Natural Science Foundation of China under grants 12173031, 12494572, 12221003, and 12322303, the Natural Science Foundation of Fujian Province of China (No. 2022J06002), the Fundamental Research Funds for the Central Universities (No. 20720240152), the Fund of National Key Laboratory of Plasma Physics (No. 6142A04240201), the China Postdoctoral Science Foundation under grant 2024M751769, and the China Manned Space Program with grant No. CMS-CSST-2025-A13.

\clearpage

\begin{thebibliography}{99}
\bibitem[Alston et al.(2020)]{2020NatAs...4..597A} Alston, W.~N., Fabian, A.~C., Kara, E., et al.\ 2020, Nature Astronomy, 4, 597. doi:10.1038/s41550-019-1002-x
\bibitem[Armitage(2010)]{2010apf..book.....A} Armitage, P.~J.\ 2010, Astrophysics of Planet Formation, by Philip J. Armitage, 294 pp. ISBN 978-0-521-88745-8 (hardback). Cambridge, UK: Cambridge University Press, 2010.
\bibitem[Artymowicz et al.(1993)]{1993ApJ...409..592A} Artymowicz, P., Lin, D.~N.~C., \& Wampler, E.~J.\ 1993, \apj, 409, 592. doi:10.1086/172690
\bibitem[Avara et al.(2016)]{2016MNRAS.462..636A} Avara, M.~J., McKinney, J.~C., \& Reynolds, C.~S.\ 2016, \mnras, 462, 1, 636. doi:10.1093/mnras/stw1643
\bibitem[Bellovary et al.(2016)]{2016ApJ...819L..17B} Bellovary, J.~M., Mac Low, M.-M., McKernan, B., et al.\ 2016, \apjl, 819, L17. doi:10.3847/2041-8205/819/2/L17
\bibitem[Bisnovatyi-Kogan \& Blinnikov(1976)]{1976SvAL....2..191B} Bisnovatyi-Kogan, G.~S. \& Blinnikov, S.~I.\ 1976, Soviet Astronomy Letters, 2, 191. doi:10.48550/arXiv.astro-ph/0003275
\bibitem[Blandford \& Znajek(1977)]{1977MNRAS.179..433B} Blandford, R.~D. \& Znajek, R.~L.\ 1977, \mnras, 179, 433. doi:10.1093/mnras/179.3.433
\bibitem[Bower et al.(2018)]{2018ApJ...868..101B} Bower, G.~C., Broderick, A., Dexter, J., et al.\ 2018, \apj, 868, 101. doi:10.3847/1538-4357/aae983
\bibitem[Burke et al.(2021)]{2021Sci...373..789B} Burke, C.~J., Shen, Y., Blaes, O., et al.\ 2021, Science, 373, 789. doi:10.1126/science.abg9933
\bibitem[Chen et al.(2024)]{2024PhRvD.110f3003C} Chen, B., Hou, Y., Li, J., et al.\ 2024, \prd, 110, 6, 063003. doi:10.1103/PhysRevD.110.063003
\bibitem[Collin-Souffrin(1991)]{1991A&A...249..344C} Collin-Souffrin, S.\ 1991, \aap, 249, 344
\bibitem[Cranmer \& van Ballegooijen(2012)]{2012ApJ...754...92C} Cranmer, S.~R. \& van Ballegooijen, A.~A.\ 2012, \apj, 754, 92. doi:10.1088/0004-637X/754/2/92
\bibitem[Czerny et al.(2004)]{2004A&A...420....1C} Czerny, B., R{\'o}{\.z}a{\'n}ska, A., Dov{\v{c}}iak, M., et al.\ 2004, \aap, 420, 1. doi:10.1051/0004-6361:20035741
\bibitem[Dai(2019)]{2019ApJ...873L..13D} Dai, Z.~G.\ 2019, \apjl, 873, L13. doi:10.3847/2041-8213/ab0b45
\bibitem[Dungey(1961)]{1961PhRvL...6...47D} Dungey, J.~W.\ 1961, \prl, 6, 47. doi:10.1103/PhysRevLett.6.47
\bibitem[D{\"u}rmann \& Kley(2017)]{2017A&A...598A..80D} D{\"u}rmann, C. \& Kley, W.\ 2017, \aap, 598, A80. doi:10.1051/0004-6361/201629074
\bibitem[El Mellah et al.(2022)]{2022A&A...663A.169E} El Mellah, I., Cerutti, B., Crinquand, B., et al.\ 2022, \aap, 663, A169. doi:10.1051/0004-6361/202142847
\bibitem[El Mellah et al.(2023)]{2023A&A...677A..67E} El Mellah, I., Cerutti, B., \& Crinquand, B.\ 2023, \aap, 677, A67. doi:10.1051/0004-6361/202346781
\bibitem[EHT Collaboration et al.(2019)]{2019ApJ...875L...4E} Event Horizon Telescope Collaboration, Akiyama, K., Alberdi, A., et al.\ 2019, \apjl, 875, 1, L4. doi:10.3847/2041-8213/ab0e85
\bibitem[Flohic \& Eracleous(2008)]{2008ApJ...686..138F} Flohic, H.~M.~L.~G. \& Eracleous, M.\ 2008, \apj, 686, 138. doi:10.1086/590547
\bibitem[Galeev et al.(1979)]{1979ApJ...229..318G} Galeev, A.~A., Rosner, R., \& Vaiana, G.~S.\ 1979, \apj, 229, 318. doi:10.1086/156957
\bibitem[Ghosh \& Abramowicz(1997)]{1997MNRAS.292..887G} Ghosh, P. \& Abramowicz, M.~A.\ 1997, \mnras, 292, 887. doi:10.1093/mnras/292.4.887
\bibitem[Giannios(2013)]{2013MNRAS.431..355G} Giannios, D.\ 2013, \mnras, 431, 355. doi:10.1093/mnras/stt167
\bibitem[Goodman \& Uzdensky(2008)]{2008ApJ...688..555G} Goodman, J. \& Uzdensky, D.\ 2008, \apj, 688, 555. doi:10.1086/592345
\bibitem[Haardt \& Maraschi(1991)]{1991ApJ...380L..51H} Haardt, F. \& Maraschi, L.\ 1991, \apjl, 380, L51. doi:10.1086/186171
\bibitem[Hankla et al.(2020)]{2020ApJ...902...50H} Hankla, A.~M., Jiang, Y.-F., \& Armitage, P.~J.\ 2020, \apj, 902, 50. doi:10.3847/1538-4357/abb4df
\bibitem[Hesse \& Cassak(2020)]{2020JGRA..12525935H} Hesse, M. \& Cassak, P.~A.\ 2020, Journal of Geophysical Research (Space Physics), 125, e25935. doi:10.1029/2018JA025935
\bibitem[Hirose et al.(2004)]{2004ApJ...606.1083H} Hirose, S., Krolik, J.~H., De Villiers, J.-P., et al.\ 2004, \apj, 606, 1083. doi:10.1086/383184
\bibitem[Hoyle(1949)]{1949srrs.book.....H} Hoyle, F.\ 1949, Cambridge [Eng.] University Press, 1949.
\bibitem[Ivanov et al.(1999)]{1999MNRAS.307...79I} Ivanov, P.~B., Papaloizou, J.~C.~B., \& Polnarev, A.~G.\ 1999, \mnras, 307, 79. doi:10.1046/j.1365-8711.1999.02623.x
\bibitem[Kanagawa et al.(2018)]{2018ApJ...861..140K} Kanagawa, K.~D., Tanaka, H., \& Szuszkiewicz, E.\ 2018, \apj, 861, 140. doi:10.3847/1538-4357/aac8d9
\bibitem[Kelly et al.(2013)]{2013ApJ...779..187K} Kelly, B.~C., Treu, T., Malkan, M., et al.\ 2013, \apj, 779, 187. doi:10.1088/0004-637X/779/2/187
\bibitem[Komissarov \& Barkov(2009)]{2009MNRAS.397.1153K} Komissarov, S.~S. \& Barkov, M.~V.\ 2009, \mnras, 397, 1153. doi:10.1111/j.1365-2966.2009.14831.x
\bibitem[Krolik et al.(1991)]{1991ApJ...371..541K} Krolik, J.~H., Horne, K., Kallman, T.~R., et al.\ 1991, \apj, 371, 541. doi:10.1086/169918
\bibitem[Lee et al.(1999)]{1999ASPC..190..173L} Lee, H.~K., Wijers, R.~A.~M.~J., \& Brown, G.~E.\ 1999, Gamma-Ray Bursts: The First Three Minutes, 190, 173. doi:10.48550/arXiv.astro-ph/9905373
\bibitem[Li \& Cao(2008)]{2008MNRAS.387L..41L} Li, S.-L. \& Cao, X.\ 2008, \mnras, 387, L41. doi:10.1111/j.1745-3933.2008.00480.x
\bibitem[Li et al.(2017)]{2017ApJ...843...21L} Li, X., Guo, F., Li, H., et al.\ 2017, \apj, 843, 21. doi:10.3847/1538-4357/aa745e
\bibitem[Li et al.(2021)]{2021ApJ...906...52L} Li, Y.-P., Chen, Y.-X., Lin, D.~N.~C., et al.\ 2021, \apj, 906, 52. doi:10.3847/1538-4357/abc88310.1002/essoar.10504346.1
\bibitem[Liang \& Price(1977)]{1977ApJ...218..247L} Liang, E.~P.~T. \& Price, R.~H.\ 1977, \apj, 218, 247. doi:10.1086/155677
\bibitem[Lin \& Papaloizou(1986)]{1986ApJ...309..846L} Lin, D.~N.~C. \& Papaloizou, J.\ 1986, \apj, 309, 846. doi:10.1086/164653
\bibitem[Lin \& Forbes(2000)]{2000JGR...105.2375L} Lin, J. \& Forbes, T.~G.\ 2000, \jgr, 105, A2, 2375. doi:10.1029/1999JA900477
\bibitem[Liska et al.(2022)]{2022ApJ...935L...1L} Liska, M.~T.~P., Musoke, G., Tchekhovskoy, A., et al.\ 2022, \apjl, 935, 1, L1. doi:10.3847/2041-8213/ac84db
\bibitem[Liu et al.(2002)]{2002ApJ...572L.173L} Liu, B.~F., Mineshige, S., \& Shibata, K.\ 2002, \apjl, 572, L173. doi:10.1086/341877
\bibitem[Liu et al.(2016)]{2016MNRAS.462L..56L} Liu, H., Li, S.-L., Gu, M., et al.\ 2016, \mnras, 462, L56. doi:10.1093/mnrasl/slw123
\bibitem[Liu et al.(2017)]{2017NewAR..79....1L} Liu, T., Gu, W.-M., \& Zhang, B.\ 2017, \nar, 79, 1. doi:10.1016/j.newar.2017.07.001
\bibitem[Loureiro et al.(2007)]{2007PhPl...14j0703L} Loureiro, N.~F., Schekochihin, A.~A., \& Cowley, S.~C.\ 2007, Physics of Plasmas, 14, 100703. doi:10.1063/1.2783986
\bibitem[Low \& Zhang(2002)]{2002ApJ...564L..53L} Low, B.~C. \& Zhang, M.\ 2002, \apjl, 564, 1, L53. doi:10.1086/338798
\bibitem[Lusso \& Risaliti(2016)]{2016ApJ...819..154L} Lusso, E. \& Risaliti, G.\ 2016, \apj, 819, 154. doi:10.3847/0004-637X/819/2/154
\bibitem[Lyubarskii(1997)]{1997MNRAS.292..679L} Lyubarskii, Y.~E.\ 1997, \mnras, 292, 679. doi:10.1093/mnras/292.3.679
\bibitem[Lyubarsky(2009)]{2009ApJ...698.1570L} Lyubarsky, Y.\ 2009, \apj, 698, 1570. doi:10.1088/0004-637X/698/2/1570
\bibitem[Manmoto et al.(1996)]{1996ApJ...464L.135M} Manmoto, T., Takeuchi, M., Mineshige, S., et al.\ 1996, \apjl, 464, L135. doi:10.1086/310097
\bibitem[Masset \& Papaloizou(2003)]{2003ApJ...588..494M} Masset, F.~S. \& Papaloizou, J.~C.~B.\ 2003, \apj, 588, 494. doi:10.1086/373892
\bibitem[McHardy(2013)]{2013MNRAS.430L..49M} McHardy, I.~M.\ 2013, \mnras, 430, L49. doi:10.1093/mnrasl/sls048
\bibitem[McKernan et al.(2012)]{2012MNRAS.425..460M} McKernan, B., Ford, K.~E.~S., Lyra, W., et al.\ 2012, \mnras, 425, 460. doi:10.1111/j.1365-2966.2012.21486.x
\bibitem[McLaughlin et al.(2024)]{2024MNRAS.529.2877M} McLaughlin, S.~A.~J., Mullaney, J.~R., \& Littlefair, S.~P.\ 2024, \mnras, 529, 2877. doi:10.1093/mnras/stae721
\bibitem[Meusinger et al.(2010)]{2010A&A...512A...1M} Meusinger, H., Henze, M., Birkle, K., et al.\ 2010, \aap, 512, A1. doi:10.1051/0004-6361/200913526
\bibitem[Narayan et al.(2003)]{2003PASJ...55L..69N} Narayan, R., Igumenshchev, I.~V., \& Abramowicz, M.~A.\ 2003, \pasj, 55, L69. doi:10.1093/pasj/55.6.L69
\bibitem[Nathanail et al.(2020)]{2020MNRAS.495.1549N} Nathanail, A., Fromm, C.~M., Porth, O., et al.\ 2020, \mnras, 495, 1549. doi:10.1093/mnras/staa1165
\bibitem[Nathanail et al.(2022)]{2022MNRAS.513.4267N} Nathanail, A., Mpisketzis, V., Porth, O., et al.\ 2022, \mnras, 513, 4267. doi:10.1093/mnras/stac1118
\bibitem[Ni et al.(2018)]{2018ApJ...852...95N} Ni, L., Lukin, V.~S., Murphy, N.~A., et al.\ 2018, \apj, 852, 2, 95. doi:10.3847/1538-4357/aa9edb
\bibitem[Paardekooper et al.(2010)]{2010MNRAS.401.1950P} Paardekooper, S.-J., Baruteau, C., Crida, A., et al.\ 2010, \mnras, 401, 1950. doi:10.1111/j.1365-2966.2009.15782.x
\bibitem[Paardekooper et al.(2023)]{2023ASPC..534..685P} Paardekooper, S., Dong, R., Duffell, P., et al.\ 2023, ASP Conference Series, 534, 685. doi:10.48550/arXiv.2203.09595
\bibitem[Paardekooper \& Papaloizou(2009)]{2009MNRAS.394.2297P} Paardekooper, S.-J. \& Papaloizou, J.~C.~B.\ 2009, \mnras, 394, 2297. doi:10.1111/j.1365-2966.2009.14512.x
\bibitem[Pollack et al.(1996)]{1996Icar..124...62P} Pollack, J.~B., Hubickyj, O., Bodenheimer, P., et al.\ 1996, \icarus, 124, 62. doi:10.1006/icar.1996.0190
\bibitem[Ripperda et al.(2021)]{2021JPlPh..87e9012R} Ripperda, B., Mahlmann, J.~F., Chernoglazov, A., et al.\ 2021, Journal of Plasma Physics, Weak Alfv{\'e}nic turbulence in relativistic plasmas. Part 2. current sheets and dissipation, 87, 5, 905870512. doi:10.1017/S0022377821000957
\bibitem[Ripperda et al.(2022)]{2022ApJ...924L..32R} Ripperda, B., Liska, M., Chatterjee, K., et al.\ 2022, \apjl, Black Hole Flares: Ejection of Accreted Magnetic Flux through 3D Plasmoid-mediated Reconnection, 924, 2, L32. doi:10.3847/2041-8213/ac46a1
\bibitem[Scepi et al.(2024)]{2024MNRAS.527.1424S} Scepi, N., Begelman, M.~C., \& Dexter, J.\ 2024, \mnras, 527, 1, 1424. doi:10.1093/mnras/stad3299
\bibitem[Shakura \& Sunyaev(1973)]{1973A&A....24..337S} Shakura, N.~I. \& Sunyaev, R.~A.\ 1973, \aap, 24, 337
\bibitem[Sirko \& Goodman(2003)]{2003MNRAS.341..501S} Sirko, E. \& Goodman, J.\ 2003, \mnras, 341, 501. doi:10.1046/j.1365-8711.2003.06431.x
\bibitem[Sironi et al.(2025)]{2025arXiv250602101S} Sironi, L., Uzdensky, D.~A., \& Giannios, D.\ 2025, , arXiv:2506.02101. doi:10.48550/arXiv.2506.02101
\bibitem[Sironi \& Spitkovsky(2014)]{2014ApJ...783L..21S} Sironi, L. \& Spitkovsky, A.\ 2014, \apjl, 783, 1, L21. doi:10.1088/2041-8205/783/1/L21
\bibitem[Spitzer(1962)]{1962pfig.book.....S} Spitzer, L.\ 1962, Physics of Fully Ionized Gases, New York: Interscience (2nd edition), 1962
\bibitem[Sun et al.(2020)]{2020ApJ...891..178S} Sun, M., Xue, Y., Brandt, W.~N., et al.\ 2020, \apj, 891, 178. doi:10.3847/1538-4357/ab789e
\bibitem[Syer \& Clarke(1995)]{1995MNRAS.277..758S} Syer, D. \& Clarke, C.~J.\ 1995, \mnras, 277, 758. doi:10.1093/mnras/277.3.758
\bibitem[Tout \& Pringle(1992)]{1992MNRAS.259..604T} Tout, C.~A. \& Pringle, J.~E.\ 1992, \mnras, 259, 604. doi:10.1093/mnras/259.4.604
\bibitem[Uzdensky \& Goodman(2008)]{2008ApJ...682..608U} Uzdensky, D.~A. \& Goodman, J.\ 2008, \apj, 682, 608. doi:10.1086/588812
\bibitem[Uzdensky et al.(2010)]{2010PhRvL.105w5002U} Uzdensky, D.~A., Loureiro, N.~F., \& Schekochihin, A.~A.\ 2010, \prl, 105, 235002. doi:10.1103/PhysRevLett.105.235002
\bibitem[Vaughan et al.(2003)]{2003MNRAS.345.1271V} Vaughan, S., Edelson, R., Warwick, R.~S., et al.\ 2003, \mnras, 345, 1271. doi:10.1046/j.1365-2966.2003.07042.x
\bibitem[Wang et al.(2018)]{2018ApJ...868...19W} Wang, J.-S., Peng, F.-K., Wu, K., et al.\ 2018, \apj, 868, 19. doi:10.3847/1538-4357/aae531
\bibitem[Ward(1997)]{1997Icar..126..261W} Ward, W.~R.\ 1997, \icarus, 126, 261. doi:10.1006/icar.1996.5647
\bibitem[Yoon et al.(2024)]{2024GeoRL..5112126Y} Yoon, Y.~D., Moore, T.~E., Wendel, D.~E., et al.\ 2024, \grl, 51, 22, 2024GL112126. doi:10.1029/2024GL112126
\bibitem[Yuan et al.(2009)]{2009MNRAS.395.2183Y} Yuan, F., Lin, J., Wu, K., et al.\ 2009, \mnras, 395, 2183. doi:10.1111/j.1365-2966.2009.14673.x
\bibitem[Yuan et al.(2022)]{2022ApJ...924..124Y} Yuan, F., Wang, H., \& Yang, H.\ 2022, \apj, 924, 124. doi:10.3847/1538-4357/ac4714
\bibitem[Zenitani \& Hoshino(2007)]{2007ApJ...670..702Z} Zenitani, S. \& Hoshino, M.\ 2007, \apj, 670, 702. doi:10.1086/522226
\bibitem[Zhang(2017)]{2017ApJ...836L..32Z} Zhang, B.\ 2017, \apjl, 836, L32. doi:10.3847/2041-8213/aa5ded
\bibitem[Zhang et al.(2018)]{2018ApJ...862L..25Z} Zhang, H., Li, X., Guo, F., et al.\ 2018, \apjl, 862, L25. doi:10.3847/2041-8213/aad54f
\bibitem[Zhang et al.(2025)]{2025arXiv250602289Z} Zhang, L., Stone, J.~M., Mullen, P.~D., et al.\ 2025, , arXiv:2506.02289. doi:10.48550/arXiv.2506.02289
\bibitem[Zhou et al.(2024)]{2024ApJ...966L...9Z} Zhou, S., Sun, M., Liu, T., et al.\ 2024, \apjl, 966, L9. doi:10.3847/2041-8213/ad3c3f
\end{thebibliography}
\end{document}